# The radio nebula of the soft gamma repeater 1806−20

S. R. Kulkarni[1], D. A. Frail[2], N. E. Kassim[3], T. Murakami[4] & G. Vasisht[1]

[1]Division of Physics, Mathematics and Astronomy 105-24, Caltech, Pasadena CA 91125, USA

[2]National Radio Astronomy Observatory, Socorro, New Mexico 87801

[3]Remote Sensing Division, Naval Research Laboratory Code 7215, Washington, DC 20375-5351

[4]Institute for Space and Astronautical Science,

3-1-1 Yoshinodai, Sagamihara, 229 Kanagawa, Japan

**Earlier[1] we had suggested that G10.0−0.3, a non-thermal nebula, most likely an old supernova remnant (SNR), was associated with the soft gamma repeater (SGR) 1806−20. Here we present new radio images obtained at the Very Large Array (VLA) of the non-thermal radio nebula G10.0−0.3. The nebula is a plerion with a hierarchy of nested amorphous components culminating in a peak. The recent dramatic detection of an X-ray point source[2] coincident with the radio peak and a hard X-ray burst[2,3] localized to G10.0−0.3 confirms the SNR-SGR association. We propose that the SGR is an isolated pulsar with both a steady and impulsive relativistic particle wind and these two together power the nebula. We note that both this SGR and SGR 0526−66[4] are offset from the centers of their SNRs which requires high space motion of the pulsar. We suggest that some natal mechanism produces high velocity and induces SGR activity.**

The new VLA observations were obtained on Sept 14.0, 1993 UT with the VLA in the "DnC" configuration. Our previous observations[1] showed that G10.0−0.3 was an amorphous non-thermal nebula, consistent with an old SNR, but these observations left open questions about the morphology of the remnant and whether the SGR was in any responsible for the non-thermal nebular emission. Multifrequency observations were conducted at 0.3, 1.4, 4.8 and 8.4 GHz to gain a better understanding of the nebula via the usual diagnostics of continuum radio astronomy: morphology, and $\alpha$, the spectral index. Here the flux density at frequency $\nu$ is $S_\nu \propto \nu^{-\alpha}$.

At the lowest frequencies a nebula of size $9 \times 6$ arcmin$^2$ is apparent (Figure 1). The total flux in these essentially fully sampled images is 3.3 Jy and 1.3 Jy at 0.3 GHz and 1.4 GHz, respectively, giving a spectral index $\alpha = -0.6$. At 1.4 GHz a plateau of emission with a full radius of $40''$ can be seen located slightly northeast of the nominal centre of G10.0−0.3. At the higher resolution of the 4.8 and 8.4 GHz images this plateau is resolved into a compact core region with a size of a few arcseconds and complex filamentary structure. This central plateau region also has $\alpha = -0.6$, as measured between 1.4 GHz and 4.8 GHz.

The limb-brightened emission, typical of shell-type SNRs, is conspicuously absent in these images. The morphology with emission from the central regions is that of a "plerion"[5]. Plerions are a subset of SNRs and are believed to be powered by a central source of relativistic particles. The hierarchical structures culminating in a peak (Figure 1) is clearly suggestive of SGR 1806−20 being located at this peak (the "radio core"). Our preliminary assertion[6] that SGR 1806−20 was located at the peak was soon followed by a dramatic localization to this peak by the *ASCA* satellite[2]. The precise VLA position[6] of the peak enabled other multi-wavelength observations and are reported elsewhere[7,8].

The nebula, as the dump of the energetic outpourings of the SGR is a convenient calorimeter. The radio and the X-ray observations allow us, for the first time, to constrain the energetics and age of an SGR. We consider two alternative neutron star models for powering the radio nebula G10.0−0.3:

(1) An isolated radio pulsar whose spin-down luminosity is primarily via a relativistic wind. Examples include synchrotron nebula powered by young pulsars such as the Crab Nebula. Less than 10% of the SNRs are classified[9] as plerions.

(2) An accreting neutron star with radio nebulae such as the ones associated with Cir X-1[10]. In most accreting neutron star binaries the accretion power appears as X-ray emission with no substantial radio emission. However, a small and rare subgroup including Cir X-1 and SS 433[11] power a bright radio nebula. Clearly, in this group, some fraction of the accretion power is harnessed to accelerate particles to relativistic energies.

A note on terminology: we note that both types of nebulae are plerions by the morphological criterion discussed above. However, we will reserve the term plerion for nebula powered by (or strongly suspected to be powered by) pulsars and "accretion powered nebula" for nebula such as the ones around Cir X-1 or SS 433.

Neither model can fully explain all of the observations. First, G10.0−0.3 has a spectral index of −0.6, distinctly steeper than the range $\alpha = 0$ to $-0.3$, characteristic of plerions[9]. However, we note that older plerions show (unexplained) spectral steepening at centimeter wavelengths[12,13], so this objection is not fatal. An accurate spectrum of G10.0−0.3 at metre wavelengths would be most helpful. Second, at a distance of $15d_{15}$ kpc the $9\times 6$ arcmin$^2$ nebula would be $39\times 26$ pc$^2$, making it the largest known plerion. These dimensions, although large, are similar to two other old, low surface brightness plerions, Vela X and G74.9+1.2 [5].

The radio properties of G10.0−0.3 more closely resemble that of the radio nebula[10] around the X-ray binary Cir X-1. They have similar sizes, spectral indices and morphology. However, there are at least two objections to the Cir X-1 model. First, the persistent X-ray luminosity, $L_X \sim 3\times 10^{35} d_{15}^2$ erg is considerably below the Eddington luminosity. It is our observation that only those binaries with super-Eddington accretion episodes have radio nebula (P. Predehl & S. Kulkarni, in prep.) e.g. SS433, Cir X-1 and perhaps Cyg X-3. Second, these rare binaries also show dramatic radio flaring events that are related to the X-ray bursts. VLA images of the plateau/core region obtained at 8.4 GHz on 3.0 Oct 1993 UT, 4 days after the first gamma ray burst[3], show no significant ($\lesssim 10\%$) changes in the flux, as compared to the image obtained from the September data. Third, at the radio peak there is no bright IR source[7] as is found towards Cir X-1, Cyg X-3 or SS433.

To summarize, the radio spectral index and the nested morphology of G10.0−0.3 favor a Cir X-1 type of model whereas the plerion model is favored by the X-ray data and the absence of radio variability. Clearly a hybrid model is required to satisfy both these observations. We propose that the persistent luminosity is due to a Vela-like pulsar which, for some mysterious reasons, undergoes super-Eddington bursts of particles. The particles are not injected in a spherical wind, as in a pulsar, but in collimated outflows or jets, as in sources like SS433 or Cir X-1. Indeed a hint of such a flow is visible in the 8-GHz image (Figure 1). If our conjecture is correct, then in the past the time averaged burst particle luminosity must have been much greater than the persistent particle luminosity. (At the present time, the time averaged burst photon luminosity appears to be less than the persistent photon luminosity). Thus in our model, the burst mechanism is not an insignificant drain on the energy budget of the pulsar.

Applying the usual synchrotron emission formulae[14] and assuming low and high frequency cutoffs of 10 MHz and 100 GHz, we estimate a minimum energy in particles and magnetic field to be $E_0 \sim 7\times 10^{49} d_{15}^{17/7}$ erg. This minimum energy is slightly higher than that for most[15] plerions but not unusually so. Equating $E_0$ to the initial rotational energy of the pulsar, $2\pi^2 I P_i^{-2}$ we infer the initial period, $P_i \lesssim 16$ ms; here $I$ is the moment of inertia. Given the *ASCA* Crab-like spectrum (i.e. $\alpha = 1$) of the X-ray source and the radio spectral index ($\alpha = -0.6$), we conclude that, as in the Crab nebula, the electromagnetic spectrum of G10.0−0.3 has a classical synchrotron spectral break ($\Delta\alpha = 0.5$) somewhere between the radio and the X-ray bands.

The *ASCA* hard X-ray source is the compact nebula formed by the shock of the relativistic wind. We predict a fainter, soft X-ray extended nebula. As with other plerions, we assume[16] a rough inefficiency of $\eta \sim 10^{-2}\eta_{-2}$ between the X-ray emission and $\dot{E}$ the spin-down luminosity of the pulsar. Thus $\dot{E} \sim 3 \times 10^{37} \eta_{-2}^{-1} d_{15}^2$ erg s$^{-1}$, leading to a characteristic nebular age, $t_n \sim E_0/\dot{E} \sim 8 \times 10^4 \eta_{-2} d_{15}^{3/7}$ yr. Since adiabatic losses could have been considerable we note that $t_n$ is an underestimate.

The 2-arcmin offset location of the pulsar with respect to the centroid of G10.0−0.3 provides a lower limit to $t_p$, the age of the pulsar. Assuming that the pulsar has a velocity of $v < 10^3$ km s$^{-1}$, the highest measured pulsar velocity[17], we obtain $t_p \gtrsim 6 \times 10^3$ yr. $t_p$ cannot be much larger than this value since it is well known that plerions fade quite rapidly[18], $t_p \lesssim 10^4$ yr. Thus the pulsar has a large velocity, $v \gtrsim 500$ km s$^{-1}$. Since $t_n >> t_p$ we conclude significant $\dot{E}$ evolution has taken place. Assuming a simple magnetic dipole spin-down model we constrain the magnetic field strength, $B \lesssim 10^{13}$ G; here we have made the reasonable assumption that the current period is significantly larger than $P_i$.

The properties of the hypothesized pulsar are not different from those of a middle-aged pulsar like the Vela pulsar. However, SGRs are very rare[19]. Thus we are forced to the conclusion that only pulsars with a special property evolve to SGRs. In this context, it is noteworthy that both 1806−20 and SGR 0526−66 (associated[4] with SNR N49) are offset from the centres of their SNRs. Thus both neutron stars have substantial space motion, $v \gtrsim 500$ km s$^{-1}$. Seemingly a connection exists between the SGR phenomenon and high velocity. We speculate that these two are manifestations of a more fundamental process. For example, models[20] of SGRs which advocate peculiar magnetic field configurations or extraordinary field strengths ($B \sim 10^{15}$ G) can conceivably generate large peculiar velocities, either at birth or shortly thereafter. Massive internal changes could be effected when the field tries to reach a more regular distribution or reconnects. These changes may then give rise to the soft gamma ray bursts. In any case, a successful model for SGRs must explain why SGRs have high velocities.

**ACKNOWLEDGEMENT.** We are indebted to M. Rupen for assistance with the October observations at the VLA. SRK gratefully acknowledges discussion with R. Blandford. DAF thanks R. Hjellming for useful discussions. SRK's work is supported by grants from NASA, NSF and the

Packard Foundation. The VLA is operated by Associated Universities Inc. under a cooperative agreement with the National Science Foundation. Basic research in Radio Interferometry at the Naval Research Laboratory is supported by the Office of Naval Research through funding document N00014-93-WX-35012, under NRL work unit 2567.

**Figure 1.** Radio images of the supernova remnant G10.0−0.3 at 1.4 GHz and 8.4 GHz. Both images were made at the VLA in the DnC configuration each with approximately 35 minutes of integration time. The 1.4 GHz image (left) has a resolution of 31″, while the 8.4 GHz image (right) has a resolution of 5.5″. Contour intervals for the 1.4 GHz image are plotted at −3, 2, 4, 6, 8, 10, 12, 14, 20, 25 and 30 times the rms noise level of 1.56 mJy beam$^{-1}$, with the negative contour indicated by dashed lines. Contour intervals for the 8.4 GHz image are plotted at −3, 2, 3, 5, 7, 9, 11, 15, 20, 25 and 30 times the rms noise level of 0.057 mJy beam$^{-1}$. A circle centred on the steady *ASCA* X-ray source[2] is shown in the 1.4 GHz; the radius of 1-arcmin is the estimated uncertainty of the X-ray source. Note that VLA images obtained in a given configuration display a restricted range of angular scales, from $\lambda/B_{max}$ to $\lambda/B_{min}$ where $B_{max}$ and $B_{min}$ are the maximum and minimum antenna separations; for the observations reported here, $B_{min}$ = 35 m and $B_{max}$ = 1900 m. The low frequency images have large field of view and are good for obtaining the overall view of the source; the highest frequency images are suited to the study of compact high surface brightness features. Bearing these caveats in mind, the morphology of the new images can be readily interpreted in terms of a compact source which is powering the entire nebula and is also the SGR. The peak of the core is $\alpha$= 18h 05m 41.7s and $\delta$ = −20° 25' 12″ (equinox 1950) and this coincides with the steady *ASCA* X-ray source.